\documentclass{article}
\usepackage[dvips]{graphicx}
\begin{document}
\title{\bf MOA-cam3: a wide-field mosaic CCD camera for a gravitational microlensing survey in New Zealand}

\author{T.\ Sako       $^{1}$,
        T.\ Sekiguchi  $^{1}$,
        M.\ Sasaki     $^{1}$,
        K.\ Okajima    $^{1}$,
        F.\ Abe        $^{1}$,
        I.\ A.\ Bond    $^{2}$,\\
        J.\ B.\ Hearnshaw $^{3}$,
        Y.\ Itow       $^{1}$,
        K.\ Kamiya     $^{1}$,
        P.\ M.\ Kilmartin $^{3}$,
        K.\ Masuda     $^{1}$,
        Y.\ Matsubara  $^{1}$,\\
        Y.\ Muraki     $^{1}$,
        N.\ J.\ Rattenbury $^{4}$,
        D.\ J.\ Sullivan $^{5}$,
        T.\ Sumi       $^{1}$,
        P.\ Tristram   $^{3}$,\\
        T.\ Yanagisawa $^{6, 7}$ and
        P.\ C.\ M.\ Yock $^{8}$}
\date{}
\maketitle

\vskip 50mm
\noindent
1. Solar-Terrestrial Environment Laboratory, Nagoya University, Nagoya 464-8601, Japan\\
2. Institute for Information and Mathematical Sciences, Massey University, Auckland, New Zealand\\
3. Department of Physics and Astronomy, University of Canterbury, Christchurch, New Zealand\\
4. Jodrell Bank Observatory, The University of Manchester, Macclesfield, UK\\
5. School of Chemical and Physical Sciences, Victoria University, Wellington, New Zealand\\
6. Advanced Space Technology Research Group, Institute of Aerospace Technology, Japan Aerospace Exploration Agency(JAXA), Japan\\
7. Orbital Debris Program Office, National Aeronautics and Space Administration(NASA), USA\\
8. Department of Physics, University of Auckland, Auckland, New Zealand

\clearpage

\begin{Large}
\begin{center}
{\bf Abstract}
\end{center}

\noindent We have developed a wide-field mosaic CCD camera,
MOA-cam3, mounted at the prime focus of the Microlensing
Observations in Astrophysics (MOA) 1.8-m telescope. The camera
consists of ten E2V CCD4482 chips, each having 2k$\times$4k
pixels, and covers a 2.2\,deg$^{2}$ field of view with a single
exposure. The optical system is well optimized to realize uniform
image quality over this wide field. The chips are constantly
cooled by a cryocooler at $-80^{\circ}$\,C, at which temperature
dark current noise is negligible for a typical 1-3 minute
exposure. The CCD output charge is converted to a 16-bit digital
signal by the GenIII system (Astronomical Research Cameras Inc.)
and readout is within 25 seconds. Readout noise of 2--3\,ADU (rms)
is also negligible. We prepared a wide-band red filter for an
effective microlensing survey and also Bessell V, I filters for
standard astronomical studies. Microlensing studies have entered
into a new era, which requires more statistics, and more rapid
alerts to catch exotic light curves. Our new system is a powerful
tool to realize both these requirements.


\clearpage

\section{Introduction} \label{INTRO}

A decade after the first detection of gravitational microlensing
events by the MACHO, EROS and OGLE groups
\cite{MACHO}\cite{EROS}\cite{OGLE}, the study of microlensing has
rapidly developed. The galactic dark matter problem which first
motivated us is well summarized by the MACHO group \cite{MACHO57}.
They concluded MAssive Compact Halo Objects (MACHOs) consists of
only 20\% of galactic dark matter. However, with only 17 events,
the statistics are not sufficient to make a definitive conclusion.
Also the nature of MACHOs, where they exist and what they are, is
not yet clearly understood.

To answer these fundamental questions, two new major survey
projects have been started. One is OGLE-III \cite{OGLE3} in Chile
and the other is our project, MOA2, in New Zealand. Each group has
a dedicated telescope and a wide-field camera to find more
microlensing events. At the same time, the development of
Difference Image Analysis (DIA) \cite{DIAORG}\cite{DIAIAN} enables
us to make more rapid (almost in real time) event detections from
huge quantities of image data.

Recently, extra-solar planet searches have become one of the main
targets of microlensing studies. Contrasting with other planet
searching techniques, the microlensing method is 
sensitive to Earth-size planets located within the habitable zone,
even with current techniques \cite{PLANETSEARCH}. This was at
first thought to be difficult, because we must catch very small
deviations from standard microlensing light curves. However,
world-wide collaboration by  survey teams (MOA, OGLE) with a rapid
alert system (using DIA) and rapid follow-up teams (PLANET,
$\mu$FUN) have recently proved that this technique is really
possible \cite{JOVIANMASS}\cite{NEPTUNE}.

Our new project, MOA2, has finished the installation of the new
1.8-m MOA telescope at the end of 2004 at Mt John University
Observatory in New Zealand. After pilot observations and tuning in
2005, we started a regular survey program and alert system in
2006. In this paper, we describe details of the wide-field camera
mounted at the prime focus of the new telescope. Details of the
project and of the telescope will be presented elsewhere. We first
describe the outline of the MOA project in Section-\ref{MOA} and
details of the camera are explained in Section-\ref{MOA-cam3}. The
actual performance of the camera is presented in
Section-\ref{PERFORM}.

\section{The MOA project (MOA and MOA2)} \label{MOA}

\subsection{MOA} \label{MOA1}

The MOA project \cite{MOAWEB} started in 1995 to search for
microlensing events occurring in the Large Magellanic Cloud (LMC)
and the Galactic Bulge (GB) directions. The team is a
collaboration of Japan and New Zealand scientists. Observations
are performed at Mt John Observatory in New Zealand
(170.5$^{\circ}$E, 44.0$^{\circ}$S, 1029~m a.s.l.) using the 61-cm
B\&C telescope at the f/6.25 Cassegrain focus. At first, we
started with a nine-CCD mosaic camera, MOA-cam1, each chip having
1k$\times$1k pixels. In 1998, the camera was upgraded to
MOA-cam2 which consists of three 2k$\times$4k CCD SITe ST-002A
chips. The details of MOA-cam2 are described by Yanagisawa et al.\
\cite{MOACAM2}

Mainly using MOA-cam2, MOA has successfully detected microlensing
events \cite{SUMI}, and also from the mass photometry database,
MOA has studied variable stars \cite{NODA}\cite{ABE}. MOA has also
succeeded in establishing a rapid alert system for microlensing
candidates \cite{DIAIAN}\cite{MOAALERT}. This enables follow-up
teams access to extensive photometry for the  most interesting
events.

\subsection{MOA2} \label{MOA2}

Following successful results from MOA1, MOA2 has been launched.
MOA2 has its dedicated telescope specially designed for a
wide-field microlensing survey. The mirror is 1.8\,m in diameter
and has a fast focal ratio value of f/3. Corrector lenses have
been designed to realize good image quality over the wide focal
plane at prime focus. The point spread caused by the optical
system is small enough over the entire focal plane, when compared
with the typical seeing at Mt John. The combined optical system of
primary mirror plus four corrector lenses results in an f/2.91
focal ratio and the available diameter of the focal plane
corresponds to 2$^{\circ}$.

MOA-cam3 is a specially designed camera installed at the prime
focus of the MOA2 telescope. It is required that the wide focal
area is covered by a well aligned mosaic of CCDs and a wide
dynamic range to perform the photometry of millions of stars.
These are crucial for the new microlensing survey programs.

The MOA2 telescope was constructed at the end of 2004. After pilot
observations and tuning in 2005, regular observations have been
performed since 2006.

\section{MOA2 prime focus camera : MOA-cam3} \label{MOA-cam3}

\subsection{Overview} \label{MOA-cam3-ov}

As described in the previous section, the optical system makes a
uniform image quality within a diameter of 2$^{\circ}$ at the
focal plane. This area is covered by ten CCD chips aligned in a $5
\times 2$ array. The CCD chips are mounted on an aluminum-nitride
ceramic plate (hereinafter AlN) and cooled down to
$-80^{\circ}$\,C (190\,K) by a cryocooler through a copper braid.
The clock, bias and output signals to/from the CCD are fed through
hermetic Dsub connectors attached to the dewar wall. Electronic
components are mounted just beside the dewar, so that even the
longest signal cable becomes less than 90\,cm. Communication
from/to the host computer, placed in the adjacent observation
room, is made through optical fibers.

All the components described above are packed into a compact
volume as shown in Fig.\ref{FIG-CAMERA-OV} and are mounted behind
the corrector lenses, {\it i.e}., at the prime focus. 
The camera is electrically insulated from the telescope body by a bakelite
plate to avoid motor driver noise from the telescope. 
Also, all the cables connected to the camera are electrically insulated 
from the telescope. 
A photo of the camera taken before mounting on the telescope is shown 
in Fig.\ref{FIG-PHOTO}

In this section, we first introduce each component of the camera.
The total performance is described in Section-\ref{PERFORM}.

\subsection{Dewar} \label{DEWAR}

The vacuum dewar is made of aluminum, except for the part of the
bellows weld. The 8\,mm thick top window is anti-reflection
coated. (Here we define the 'top' as the direction from which the
photons come.) Its transmissivity is more than 95\% over a wide
wave band, 400-950~nm.

Ten hermetic 37-pin Dsub connectors and one hermetic 40-pin round
connector are attached on the side wall. Each Dsub connector
corresponds to one CCD chip. The round connector is intended for
multi-purpose use. This is actually used for monitoring the
temperatures at the cryocooler cold head and the AlN plate. A
heater is attached on the cooling path. Current for the heater can
be fed through this connector when necessary.

The cryocooler cold head is also inserted in the vacuum space. To
avoid the transfer of any vibration to the CCD chips, a bellows is
placed between the main dewar and the cold head. The cryocooler is
supported by four poles, which are connected to the heavy image
rotator, that is a part of the telescope. Any cryocooler vibrations are
damped out through these poles.

\subsection{Vacuum and cooling system} \label{VACUUMCOOL}

A Gifford-McMahon cycle cryocooler GR-101 (AISIN SEIKI Co., Ltd.)
is used to cool down the CCD chips. The weight of the cold head
unit mounted on the camera is 5\,kg. Compressed helium gas is fed
through a 15-m flexible hose from a compressor located on the
turn-table of the telescope. The cryocooler has 12\,W cooling
power and is connected to a 100\,V/50\,Hz power supply. A naive
estimation of the heat injection to the camera is about 11\,W,
which is almost all thermal radiation from the window and wall.
(Details of the calculation method are found in
\cite{SUPRIME-CAM}.) Because we know this value is an overestimate
for the real case, a cooling power of 12\,W is deemed sufficient.
The actual achievable temperature depends on the structure, such
as the copper braid cross section, surface contact of each cooling
component, vacuum level, outside temperature and so on. As the
power of the cryocooler is not adjustable, we must operate it
always at maximum power. The target is to achieve $-80^{\circ}$\,C
when the outside temperature is 10$^{\circ}$\,C, which is the
usual temperature during the summer nights at Mt John. Because we
expect some level of over-cooling, a 3\,W resistance heater is
placed below the CCD mounting plate. By controlling the current in
this resistance, we can keep a constant plate temperature.

The cold head of the cryocooler has its own vibrational motion
during operation. The horizontal and the vertical oscillations of
the cold head itself have been directly measured. These are shown
in Fig.\ref{FIG-CH-VIB}. As shown in the figure, the vertical
oscillation peak-to-valley amplitude is 14\,$\mu$m, which is less
than the focal depth of 45\,$\mu$m. The horizontal oscillation
peak-to-valley amplitude is 7\,$\mu$m, which corresponds to half a
pixel size and also to 0.3\,arcsec. This is almost negligible when
compared with the typical seeing of 2.0~arcsec at Mt John.
Furthermore, we must note that the cold head vibration does not
directly transfer to the image plane, because the cold head is
connected to the AlN plate via a copper braid.

The cooling power and the vacuum level are strongly correlated.
When the vacuum level is worse (that is the pressure is
higher), thermal conduction results in more heat input and the
temperature rises. When the temperature becomes higher, molecules
adsorbed on the cold components are released, and the pressure
increases. Above a certain critical level, this positive feedback
catastrophically increases the CCD temperature. In the case of our
camera, the critical level was roughly around 10$^{-3}$ mbar in
pressure. So we tried to keep below this pressure after removing
the dewar from the vacuum pump. The pressure in the dewar and
temperatures at the AlN and the cold-head are always monitored by
a remote PC. Fig.\ref{FIG-VAC-T} shows long-term temperature and
pressure profiles measured at Mt John in 2005 August. The curves
show stable temperature and pressure profiles for a month.
The AlN temperature is well maintained around our target temperature of 
$-80^{\circ}$\,C. 
From the laboratory tests, we know that the thermal noise at 
$-70^{\circ}$\,C is typically 1~ADU in 100~s, and it starts to increase 
when the temperature exceeds $-60^{\circ}$\,C. We can see that our
observations are all made below this limit.
As no trend is found during a month, we don't expect to have to
vacuum pump the dewar as frequently as every month. However, in
the New Zealand summer time, we need pumping every few days to a week. 

The diurnal variation of $\Delta$T$\sim30$\,K causes a scale
variation of AlN to be 24\,$\mu$m (1.6 pixel or 0.93\,arcsec) over the
field of view (200\,mm in diameter).
Because our analysis is performed dividing the field into small
subareas of 1k$\times$1k pixels, the scale variation is reduced to be
0.1 pixel that can be compensated by DIA.
Some investigations to keep the temperature more stable are now under progress.
Improving the dome temperature control in summer time and driving the
cryocooler with a higher frequency (60\,Hz) should keep the
temperature below $-80^{\circ}$\,C all the time.
Installing an ion pump may also be a promising way, but we must work within
a limited volume at the prime focus.
Once the camera is overcooled, we can control the temperature by using
the heater installed in the camera so that we can keep a constant temperature.

As a reference, we also show two other profiles in
Fig.\ref{FIG-SIEVES}. These plots show the pressure evolution
after removing the dewar from the pump. The only difference
between the two plots is whether or not use molecular sieves in the dewar. 
In the actual setting, we attached 100~g of Molecular Sieves 5A 1/16 (WAKO) 
at the cold head, where it is usually cooled down to 
$-150^{\circ}$\,C (120\,K).
Because the adsorption of the molecular sieves is more effective at low
temperature, they are packed in a copper box with a copper mesh at one surface.

\subsection{CCD} \label{CCD}

MOA-cam3 uses ten E2V CCD4482 chips. Each chip has 2k$\times$4k
pixels and is a  back-illuminated CCD. The pixel size is
15\,$\mu$m square, which corresponds to 0.58\,arcsec when combined
with the f/2.91 optics. 
The cataloged sensitivity and the full dynamic range are 6.0\,$\mu$V/electron
and 150k-200k electrons, respectively. 
The quantum efficiency as a function of wavelength is shown in
Fig.\ref{FIG-CCDQE}.

We arranged 10 chips in a $5 \times 2$ array resulting in a
10k$\times$8k mosaic camera for the total system. The gap between
chips is set to be 0.5\,mm. Because each chip has an insensitive
area of 0.5\,mm on both sides and 0.17\,mm at the top (opposite to
the readout side), the inter-chip inactive space becomes 1.5\,mm
and 0.84\,mm, respectively. Excluding these inactive areas, the
chips cover 2.18\,deg$^{2}$.

The chips must maintain a flatness of within $\pm$45\,$\mu$m,
which results in one pixel size of defocussing. We prepared an AlN
mother plate to mount the chips. Because the thermal expansion
coefficient of AlN, 4.0$\times$10$^{-6}$\,K$^{-1}$, is very
similar to that of the package of the chips,
1.2$\times$10$^{-6}$\,K$^{-1}$, which is made of invar, even a
large temperature change of $\Delta T \sim 120$\,K, does not
damage the chips. At the same time, the thermal conductivity of
AlN, 170\,W/m/K, is high, so that effective cooling can be
realized. The flatness of the AlN plate is kept to within
$\pm$10\,$\mu$m (min-max). Also the flatness of each chip is
guaranteed to be within $\pm$20\,$\mu$m (min-max). After mounting
the chips on the plate, we measured the flatness of the whole chip
array. The relative depth of a total 832 points on the chips was
measured using an automatic non-contact 3-dimensional measurement
stage (NH-6, Mitaka Kohki Co., Ltd.) at Nagoya University. The
result is shown in Fig.\ref{FIG-CCDSURFACE} and
Fig.\ref{FIG-CCDSURFACEZ}. As seen in the figure, the depth
variation is within 25\,$\mu$m (min-max). We can even find the
surface structure of each chip. This is a satisfactory result
which conforms to the optical requirements of the telescope and
detector system.

The AlN plate is supported on the dewar via four polycarbonate
poles. Because the thermal conductivity of polycarbonate is
0.1\,W/m/K, the expected heat injection from these supports is
only 0.08~W, so that the conducted heat input is negligible.

Fig.\ref{FIG-ARRAY} is a photo of the CCD array mounted on the AlN plate
and placed in the camera dewar.

\subsection{Electronics} \label{ELEC}

To drive an E2V CCD4482 chip, we need nine clock signals and five
DC biases. To feed these signals to the chips, we used the GenIII
system from Astronomical Research Cameras Inc.\ (hereinafter ARC).
The system contains one timing board which communicates with the
host computer via optical fiber cables and controls all the other
boards. Two clock boards are used to provide clock signals to the
chips. Each board sends an identical set of clock signals to five
chips. Five video boards are used to feed DC biases, which are
tuned for each chip. Each video board can generate 12 independent
DC biases, and they are separated to two chips. A video board also
has two 16-bit A/D converters to digitize the output signal from
two chips. The A/D converters follow the amplifiers, and the
correlated double-sampling method is used to pick-up the correct
signal for each pixel.

The signal integration time, bias voltages, amplifier gain (it is
selectable from 1, 2, 5, 10) are all consistently adjusted to
match the chip and A/D dynamic ranges, which means that 200k
electrons correspond to 2$^{16}$ ADU (3.1\,e$^{-}$/ADU).
The results of readout tuning are introduced in Section-\ref{PERFORM}

The commands and clock pattern are recorded on the DSP chip on the
timing board. Observers communicate with the timing board from the
host computer, where a PCI board is installed. In our system, the
host computer is operated under Linux OS, and the camera control
commands are implemented in the C language using the libraries provided
by ARC.

\subsection{Peripherals} \label{OTHERS}

The shutter and filters are also controlled from the camera
control computer via RS232C. The host computer communicates with
another computer to control the telescope and all the commands to
the electronics, shutter, filters and telescope are synchronized.

The shutter is placed between the camera window and the last
corrector lens element. The bidirectional roll-type shutter
enables a uniform exposure over the wide image area. Filters are
located in the space between the second and third of the four
corrector lenses. They are usually parked parallel to the
telescope axis, but turn perpendicular into the light path when in
use. This system can reduce the light obscuration by the filters
when parked.
We prepared a wide-band red filter for an
effective microlensing survey and also Bessell V, I filters for
standard astronomical studies.

To avoid frosting on the camera window, nitrogen gas is introduced
into the gap between the window and last lens. Because the space
is well packed, only a slow flow of gas is sufficient.

\subsection{Performance} \label{PERFORM}

After adjusting the DC bias voltages for each chip, we have
measured the dynamic range and linearity using LED light. An
example of LED calibration for one chip is shown in
Figure-\ref{FIG-RANGE}. Figure-\ref{FIG-RANGE} a) shows the mean
ADU in a small 100$\times$100-pixel area as a function of the LED
illumination time. The deviation from linearity is only 1\% at
maximum and is limited by the stability of the LED intensity.
Figure-\ref{FIG-RANGE} b) shows a relation between the mean and
the variance of the ADU in the same small area. When the variance
is dominated by photon statistics, we can expect a linear
relation, as seen in the plot below 40\,000 ADU. Here the slope
gives the inverse of the AD conversion coefficient, the so-called
gain. The constant offset is determined by the readout noise that
is measurable from a dark image with zero exposure. 
The chip-to-chip variation of the gain and the readout noise are
2.0--2.4\,e$^{-}$/ADU and 5--7 electrons (2--3\,ADU) in rms, respectively. 
The gain values are smaller than the expected one introduced in 
Section-\ref{ELEC} because the coarse gain of the GenIII amplifier
could not be compensated.
The measured readout noise is consistent with the cataloged value of GenIII. 
We also found electrical noise due to the telescope motor drivers and reduced
it at the level of the readout noise. 
This was achieved by electrically insulating the camera and the telescope as 
described in Section-\ref{MOA-cam3-ov} while the rearrangement of grounding 
did not work as effectively as insulation.
Because the dark sky background level at Mt John in a 1-minute
exposure is typically a few 1000 ADU/pixel, these electrical sources of noise
are negligible. 
When the CCD saturates, the linearity breaks down, as is seen in 
Figure-\ref{FIG-RANGE} b) at a mean ADU of 40\,000. 
This effect is also visible as a distortion of the image. 
The breaking points are around 30\,000--40\,000 ADU in the 10 chips.
Combining with the gain, the measured dynamic range (80k electrons in average)
is half of the catalog value.
We could not find better settings whilst keeping the gain variation of ten 
chips as small as possible.

To achieve the condition described above, we chose a readout speed of 
3.1\,$\mu$s/pixel. 
Because the signal transfer is far faster than the A/D conversion, the 
readout to the host computer is done in serial for the 10 chips. 
Then the total readout for the 10 chips is completed in 25 seconds. 
These measurements and tunings were carried out both in the laboratory
and after the camera was mounted on the telescope.
Fig.\ref{FIG-LMC} shows an image of a part of the Large Magellanic
Cloud taken by the new MOA telescope and MOA-cam3. The Tarantula
nebula (30 Doradus; $\sim$0.4$^{\circ}$ in diameter), seen in the
right bottom corner, demonstrates the width of the field.

\section{Summary} \label{SUMMARY}

The mosaic CCD camera MOA-cam3 was constructed to be installed at
the new 1.8\,m MOA2 telescope prime focus. It consists of ten E2V
CCD4482 chips, each having 2k$\times$4k pixels. Combined with the
f/2.91 optics, it covers 2.2 deg$^{2}$ in a single exposure. The
chips are well aligned on the mother plate, with a satisfactory
flatness of 25\,$\mu$m peak-to-valley. The chips are cooled by a
cryocooler to around $-80^{\circ}$\,C, at which temperature the
dark current of the chips is negligible during the typical
exposure time of a few minutes. The chips are controlled from a
Linux PC via the GenIII system. The gain and dynamic range of the
chips are tuned by the DC bias voltage, and readout speed is set
to match the 16-bit ADU and chip saturation level. The readout
noise after tuning is 2--3\,ADU in rms, which is negligible in
relation to the sky background for a typical exposure. There is
also noise due to the telescope drive. 
We reduced it to the level of the readout noise by electrically 
insulating the camera and telescope body. 
The host computer also controls the shutter and filters. 
By communicating with the telescope control PC, all the observations 
are automatically processed.
Although MOA-cam3 is already in operation with sufficient power for a 
dedicated microlens survey program, some improvements described in this 
manuscript are also under progress to make it more general purpose instrument.

MOA2 started regular observations in 2006. Using a real time analysis 
technique, a rapid microlensing alert system has also been started. 
This is being upgraded to be yet more rapid and reliable in 2007. 
It means that the other follow-up teams have access to plenty of exotic 
microlensing events, which is vitally important in this new field of research. 
At the same time, our wide-field camera has an excellent capability for 
monitoring exotic microlensing events without the need for external alerts, 
as well as discovering serendipitous astronomical phenomena.

\vskip 5mm {\noindent\bf Acknowledgments}

\noindent This work is supported by a grant-in-aid for scientific research 
of the Japan Ministry of Education, Science, Sports and Culture. 
The authors are grateful to the staff of the National Astronomical Observatory 
Japan (Y.\ Kobayashi, S.\ Miyazaki, Y.\ Komiyama, H.\ Nakaya), to members of 
the Graduate School of Science (S.\ Sato, H.\ Shibai) and Instrument 
Development Group of the Technical Center of Nagoya University, to J.\ Hiraga 
in ISAS/JAXA, to staff of the Nishimura Co.\ Ltd., to H.\ Kondoh in
AISIN, to A.\ Rakich in IRL and to B.\ Leach in ARC.
Finally the authors thank the anonymous reviewer to improving the manuscript
and for general expert comments on the camera.

\clearpage

\begin{figure}
  \begin{center}
    \includegraphics[width=130mm]{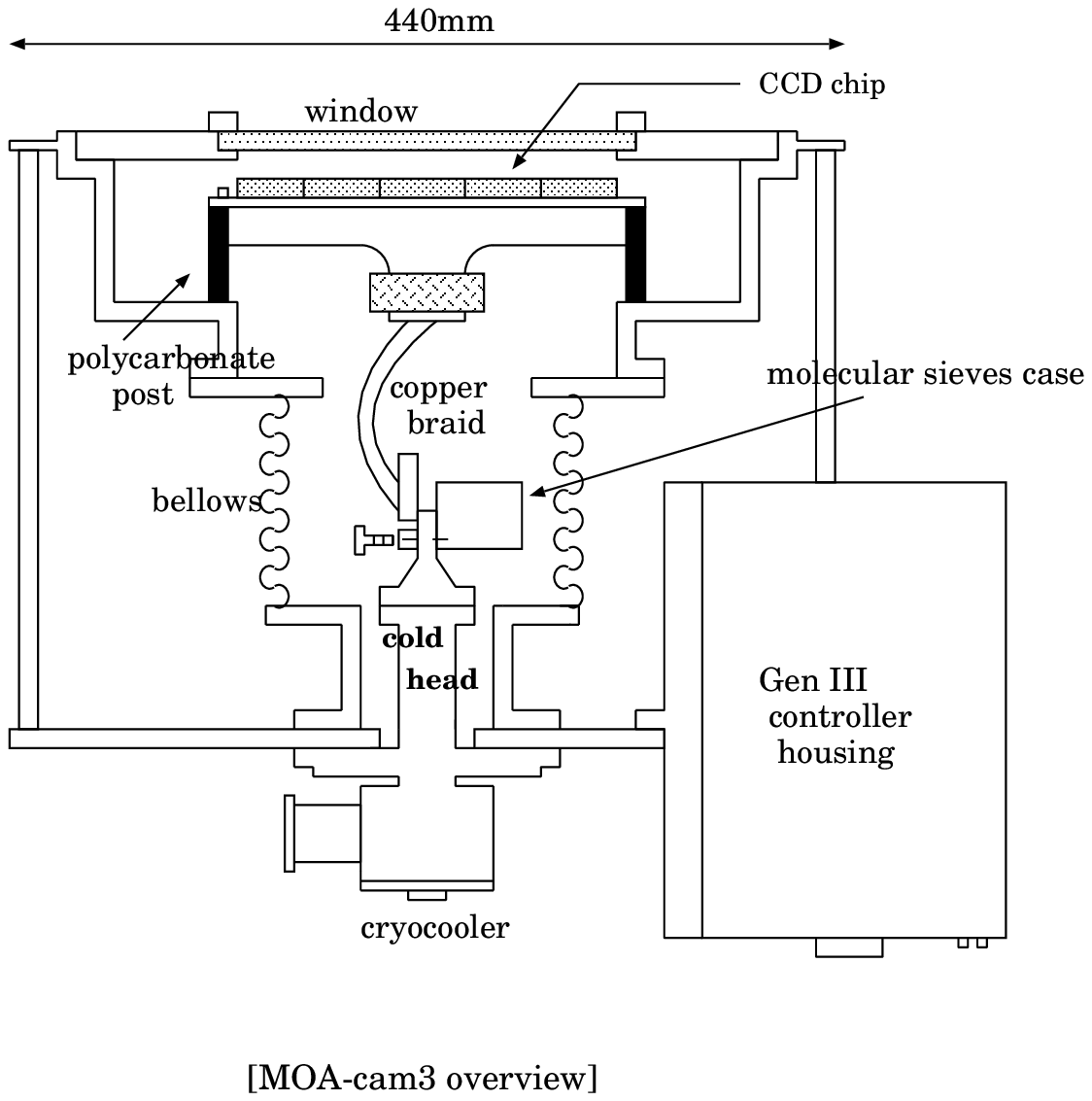}
    \caption{Schematic view of MOA-cam3.}
    \label{FIG-CAMERA-OV}
  \end{center}
\end{figure}

\begin{figure}
  \begin{center}
    \includegraphics[width=120mm]{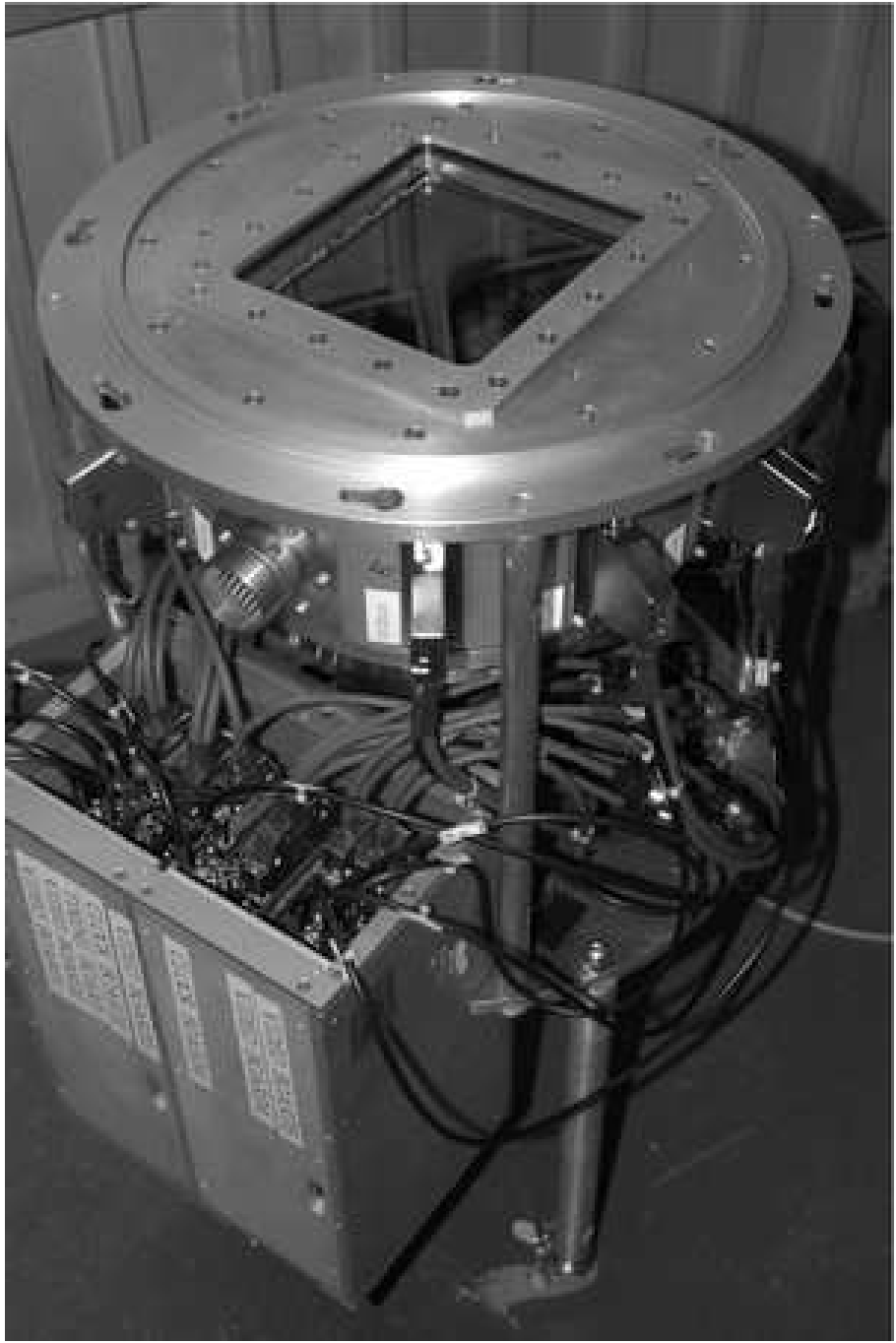}
    \caption{Photo of MOA-cam3.}
    \label{FIG-PHOTO}
  \end{center}
\end{figure}

\begin{figure}
  \begin{center}
    \includegraphics[height=110mm]{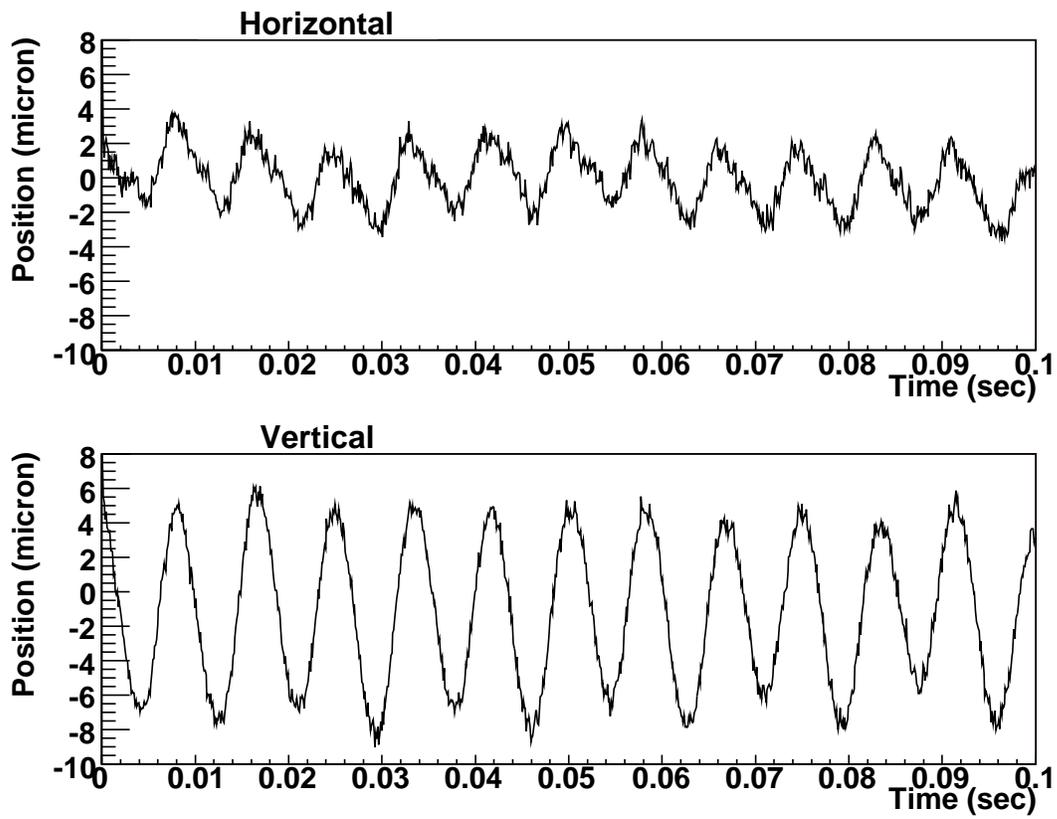}
    \caption{Vibration of the cryocooler cold head.
Relative position of the cold head during its operation was measured in the
horizontal and vertical directions.}
    \label{FIG-CH-VIB}
  \end{center}
\end{figure}

\begin{figure}
  \begin{center}
    \includegraphics[width=130mm]{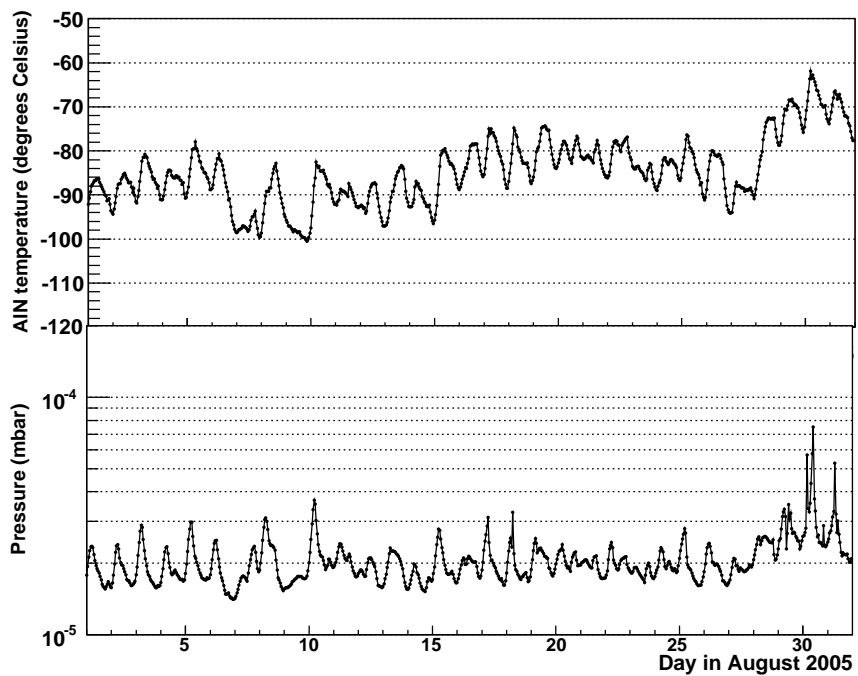}
    \caption{Long term profile of the inner dewar pressure and AlN temperature.
Though daily variation correlating to the outside temperature is seen, AlN
temperature is well kept around --80$^{\circ}$C for one month.}
    \label{FIG-VAC-T}
  \end{center}
\end{figure}

\begin{figure}
  \begin{center}
    \includegraphics[width=130mm]{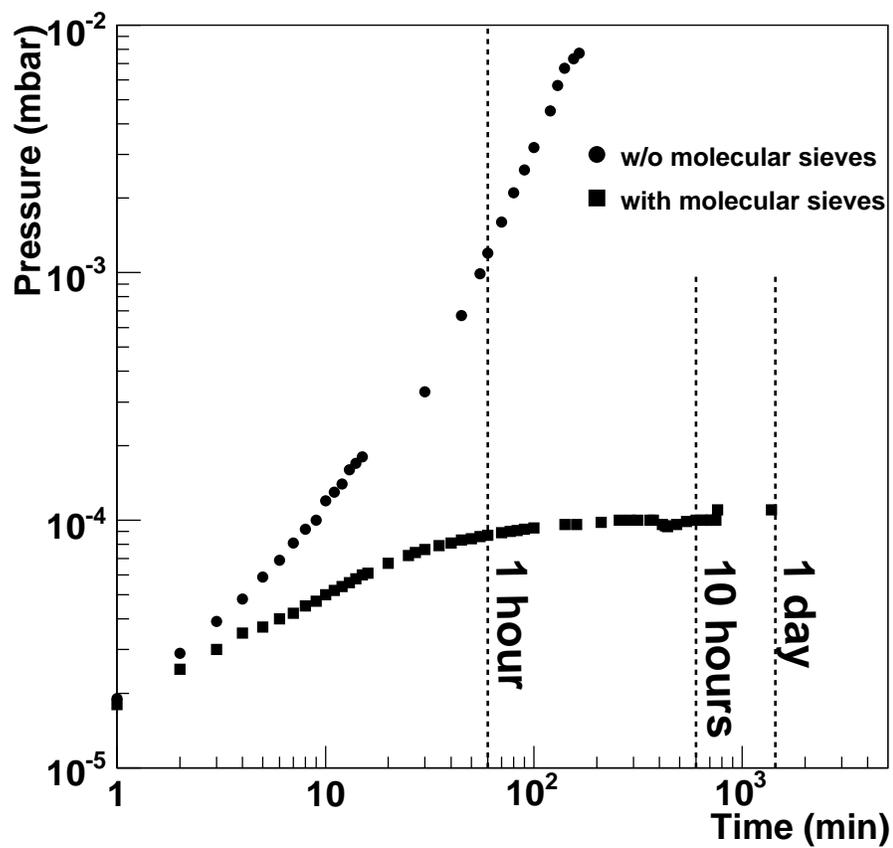}
    \caption{Time profiles of the dewar pressure with and without using molecular sieves.
The start time is defined at the moment when the valve was closed after
sufficient vacuum pumping.
The circles and the squares indicate for the case of without/with the molecular
sieves, respectively.
The effect of the molecular sieves is apparent.}
    \label{FIG-SIEVES}
  \end{center}
\end{figure}

\begin{figure}
  \begin{center}
    \includegraphics[width=130mm]{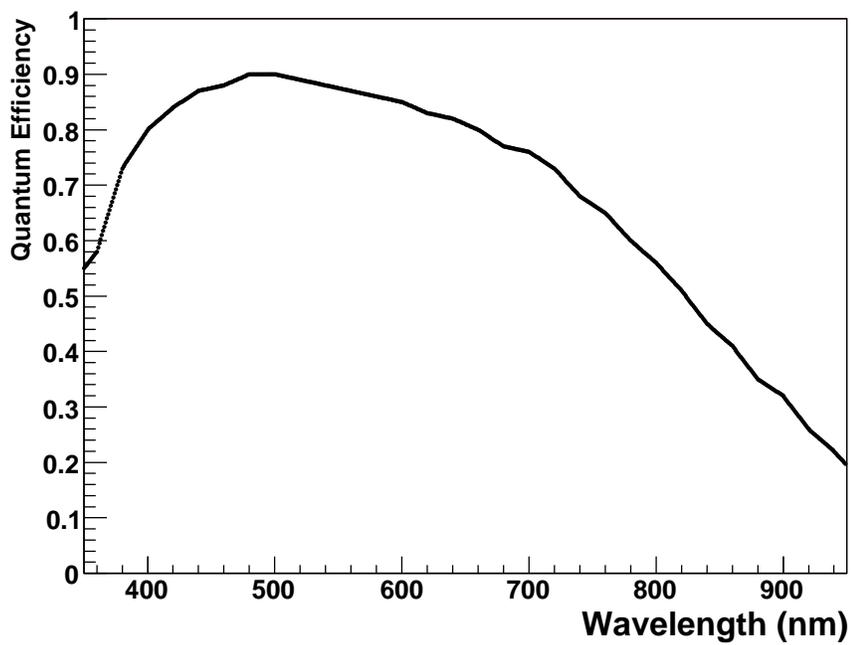}
    \caption{Quantum efficiency of E2V CCD4482.}
    \label{FIG-CCDQE}
  \end{center}
\end{figure}

\begin{figure}
  \begin{center}
    \includegraphics[width=130mm]{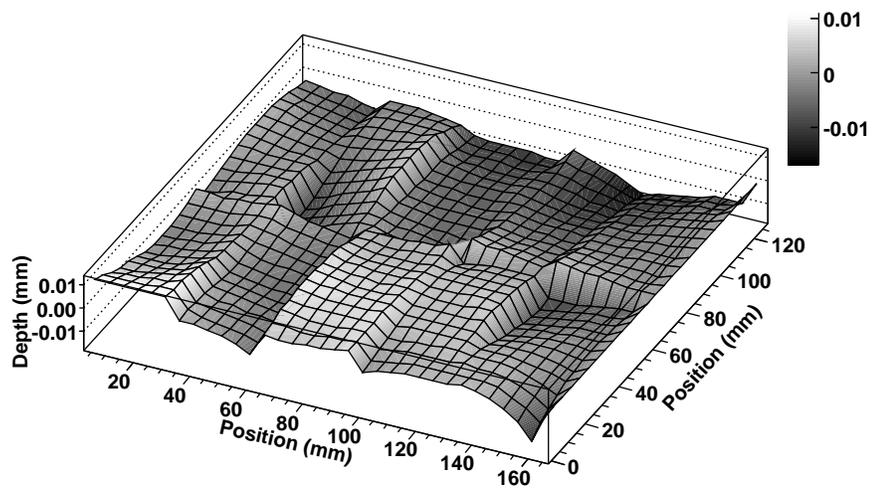}
    \caption{Surface flatness of the CCD array.
Grey scale indicates the relative depth of each measurement point.
Total 832 points were measured.}
    \label{FIG-CCDSURFACE}
  \end{center}
\end{figure}

\begin{figure}
  \begin{center}
    \includegraphics[width=130mm]{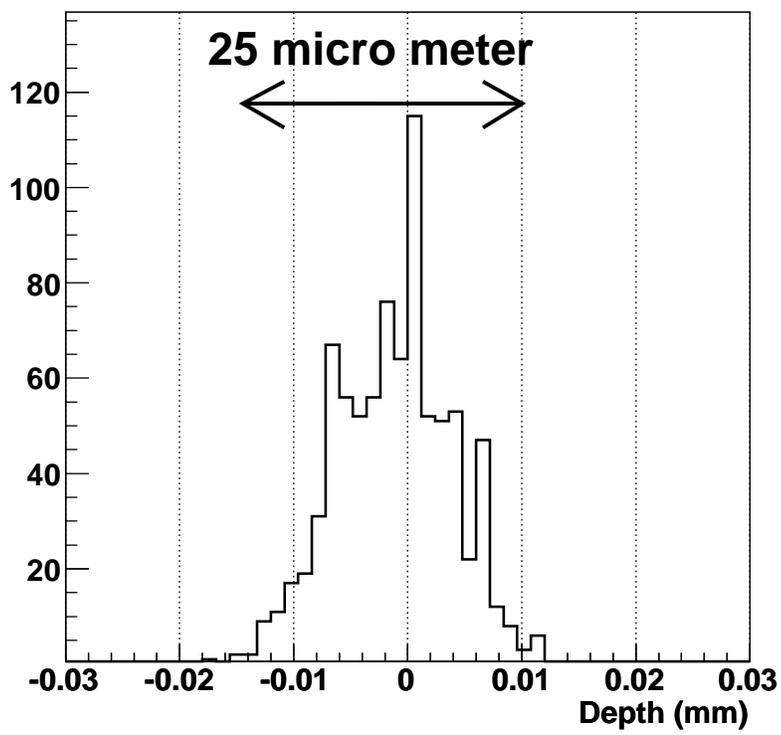}
    \caption{Depth distribution of the Fig.\ref{FIG-CCDSURFACE}.}
    \label{FIG-CCDSURFACEZ}
  \end{center}
\end{figure}

\begin{figure}
  \begin{center}
    \includegraphics[width=130mm]{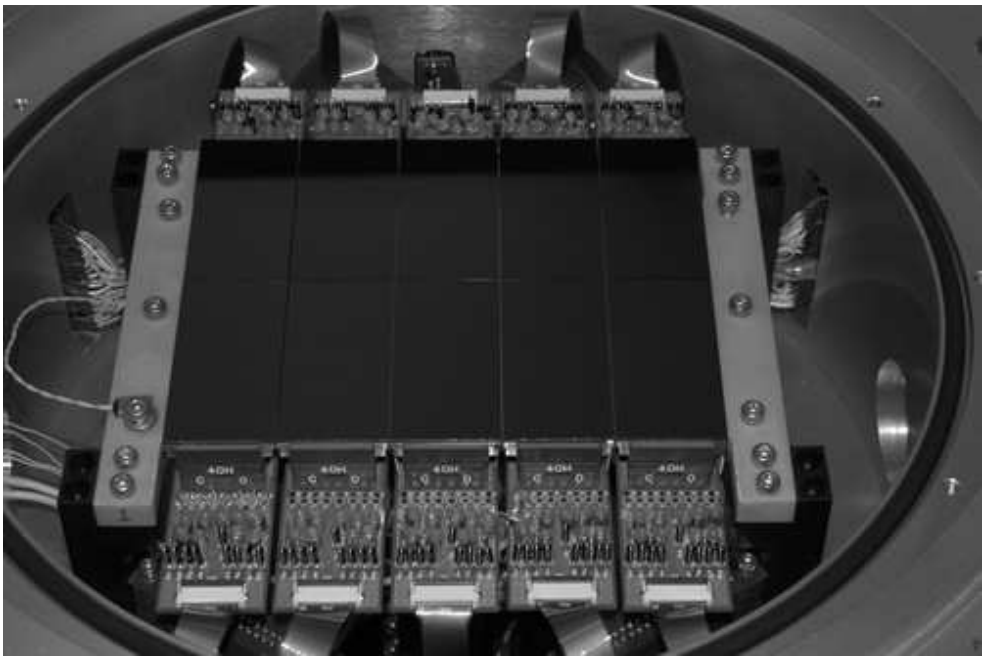}
    \caption{Photo of the ten CCD array mounted on the AlN plate.}
    \label{FIG-ARRAY}
  \end{center}
\end{figure}

\begin{figure}
  \begin{center}
    \includegraphics[width=130mm]{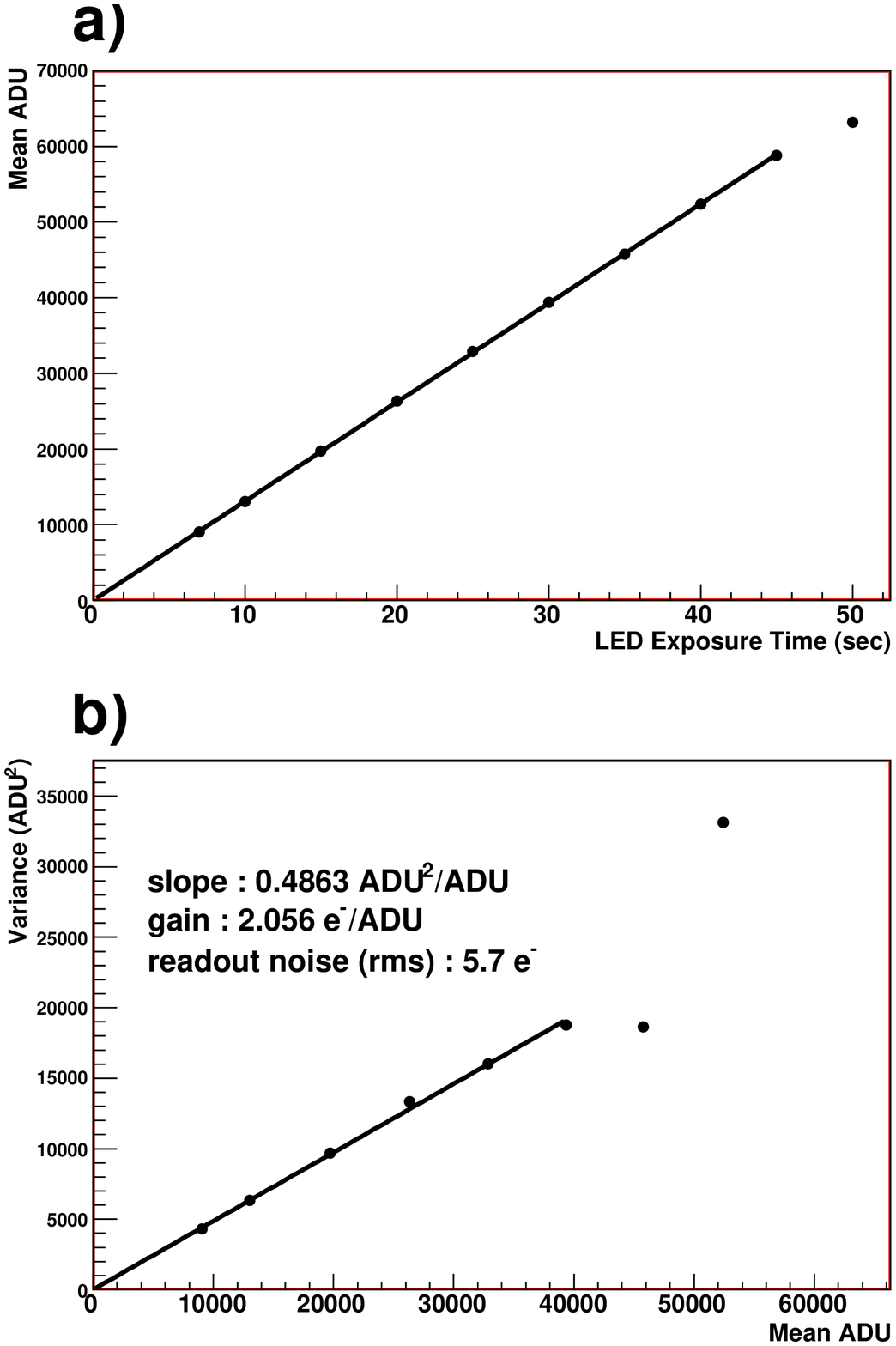}
    \caption{Measurement of the CCD linearity and dynamic range.}
    \label{FIG-RANGE}
  \end{center}
\end{figure}

\begin{figure}
  \begin{center}
    \includegraphics[width=150mm]{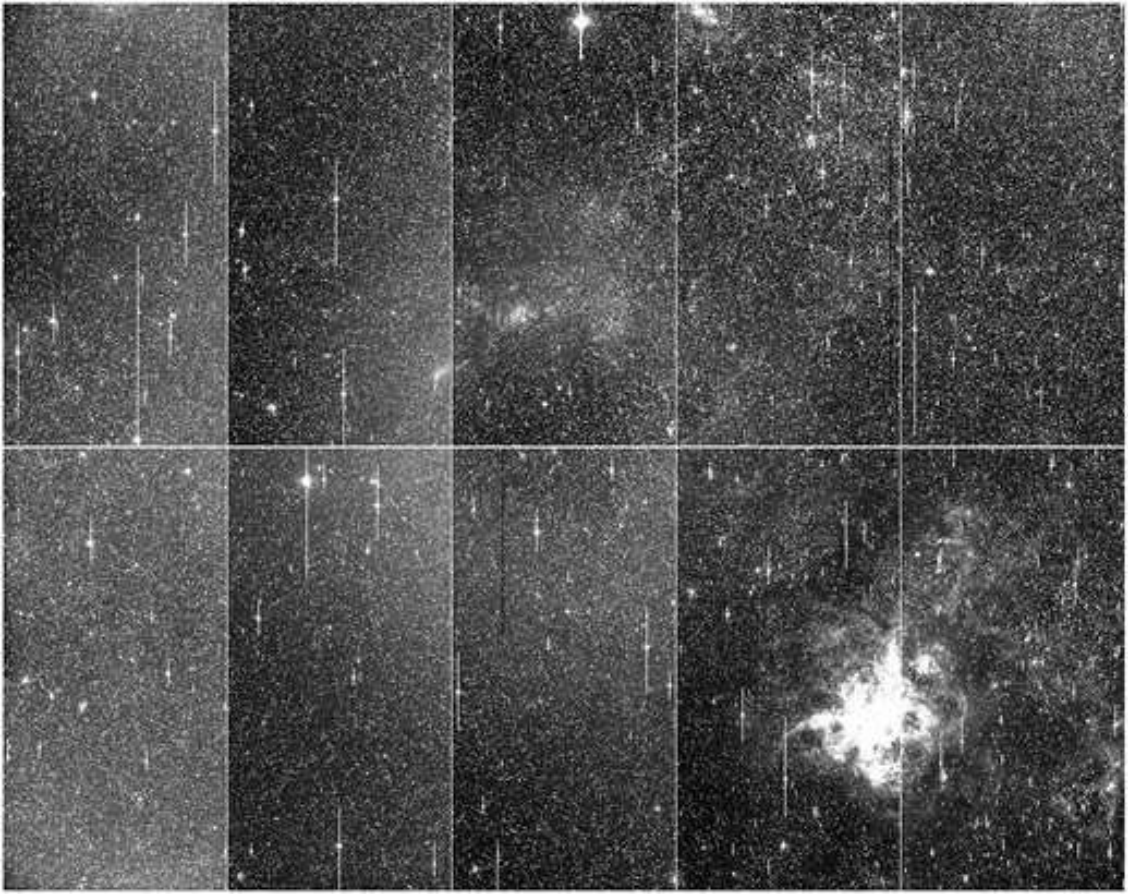}
    \caption{Combined 10 chip image of a part of the Large Magellanic Cloud taken by MOA-cam3.
The Tarantula nebula (30 Doradus; $\sim$0.4$^{\circ}$ in diameter) seen in the
right bottom corner shows the wide field of the system.
}
    \label{FIG-LMC}
  \end{center}
\end{figure}


\end{Large}

\end{document}